# The state of protoplanetary material 10 Myr after stellar formation: circumstellar disks in the TW Hydrae association.

K. I. Uchida[1], N. Calvet[2], L. Hartmann[2], F. Kemper[3], W. J. Forrest[4], D. M. Watson[4], P. D'Alessio[5], C. H. Chen[6], E. Furlan[1], B. Sargent[4], B. R. Brandl[7], T. L. Herter[1], P. Morris[8], P. C. Myers[2], J. Najita[9], G. C. Sloan[1], D. J. Barry[1], J. Green[4], L. D. Keller[10], P. Hall[1]

kuchida@isc.astro.cornell.edu

## ABSTRACT

We have used the Spitzer Infrared Spectrograph to observe seven members of the TW Hya association, the nearest stellar association whose age ($\sim$ 10 Myr) is similar to the timescales thought to apply to planet formation and disk dissipation. Only two of the seven targets display infrared excess emission, indicating that substantial amounts of dust still exist closer to the stars than is characteristic of debris disks; however, in both objects we confirm an abrupt short-wavelength edge to the excess, as is seen in disks with cleared-out central regions. The mid-infrared excesses in the spectra of Hen 3-600 and TW Hya include crystalline silicate emission features, indicating that the grains have undergone significant thermal processing. We offer a detailed comparison between the spectra of TW Hya and Hen 3-600, and a model that corroborates the spectral shape and our previous understanding of the radial structure of these protoplanetary disks.

[1]Center for Radiophysics and Space Research, Cornell University, Space Sciences Building, Ithaca, NY 14853-6801

[2]Harvard-Smithsonian Center for Astrophysics, 60 Garden Street, Cambridge, MA 02138

[3]Spitzer Fellow, Department of Physics and Astronomy, University of California, Los Angeles, CA 90095-1562

[4]Department of Physics and Astronomy, University of Rochester, Rochester, NY 14627-0171

[5]Centro de Radioastronomia y Astrofisica, UNAM, Apartado Postal 3-72 (Xangari), 58089 Morelia, Michoacán, Mexico

[6]National Research Council Resident Research Associate, Jet Propulsion Laboratory, M/S 169-506, California Institute of Technology, 4800 Oak Grove Drive, Pasadena, CA 91109

[7]Sterrewacht Leiden, P.O. Box 9513, 2300 RA Leiden, Netherlands

[8]Spitzer Science Center/Infrared Processing and Analysis Center, California Institute of Technology, Pasadena, CA 91125

[9]National Optical Astronomy Observatory, 950 North Cherry Avenue, Tucson, AZ 85719

[10]Department of Physics, Ithaca College, Ithaca, NY 14850

*Subject headings:* accretion, accretion disks, stars: circumstellar matter, stars: formation, stars: pre-main sequence

## 1. Introduction

The TW Hya association (TWA) is an ideal laboratory for studies of disk evolution. Although its first members were identified more than twenty years ago (Rucinski & Krautter 1983; de La Reza et al. 1989; Gregorio-Hetem et al. 1992), it was not until the advent of *Hipparcos* that the importance of the association was fully revealed (Kastner et al. 1997). With *Hipparcos* distance measurements, stars in the association could be compared in color-magnitude diagrams with pre-main-sequence stellar models, and ages of 5 – 15 Myr (Webb et al. 1999; Weintraub et al. 2000) were inferred thereby. This is an extremely interesting age range, because a wide variety of estimates indicate that the characteristic times for disk dissipation (Briceño et al. 2001; Haisch, Lada, & Lada 2001) and planet formation (e.g. Pollack et al. (1996)) are similar.

Analysis of the spectral energy distribution (SED) of objects in TWA indicates that dust in the disks of these systems may already be actively evolving toward planetesimals and planets. Out of the ∼24 stars presently known in the association, three (TW Hya, Hen 3-600A, and HD 98800) show large infrared excesses that are similar to those of disks in younger T Tauri stars that are still surrounded by their natal molecular cloud material (Jayawardhana et al. 1999b), as well as SEDs that look similar to the median SED in Taurus at long wavelengths. These three, however, also exhibit a clear deficit of flux at $\lambda < 10\mu$m, which suggests that the dust in the inner portion of the disk may already have grown and settled to the mid-plane, as predicted by theories of dust evolution in the Solar nebula (e.g., Weidenschilling 1997).

Three low mass stars in the TWA region, TW Hya, Hen 3-600A, and TWA 14, show the unmistakable signatures of gas still accreting onto the star (Muzerolle et al. 2000; Muzerolle, Calvet, & Hartmann 2001). To support such mass flow, gas must still exist in the inner disk region. Calvet et al. (2002) have analyzed in detail the spectrum of TW Hya, showing that its disk is truncated at ∼4 AU and that matter interior to this radius carries a minute amount of small grains. They speculate that a forming planet may be responsible for the central clearing.

A few stars in the TWA region exhibit the silicate dust feature at ∼10 $\mu$m in emission. There are conflicting views as to the nature and origin of silicates. Amorphous silicates are found univer-sally in the interstellar medium and are found in young (∼1 Myr old) solar-mass objects (Natta, Meyer, & Beckwith 2000). In contrast, comets often show crystalline silicates (Hanner, Lynch, & Russell 1994; Crovisier et al. 1997), suggesting that processing took place in the early phases of the Solar nebula. The presence of crystalline features and evidence of grain growth in TWA would set the timescales for the crystallization events that occur in young disks.

Silicate emission has been reported in TW Hya (TWA 1) by Sitko, Lynch, & Russell (2000). Weinberger et al. (2002) observed structure in the silicate feature which they interpreted as due to crystalline silicates of unknown identity. The conclusive argument for crystalline silicates in Hen 3-600A was presented by Honda et al. (2003b), who found that the features of the silicate profile could be explained by a mixture of crystalline forsterite, crystalline enstatite, silica, and glassy olivine grains.

In this paper, we present Spitzer Infrared Spectrograph[1] spectra of a sample of stars in TWA. We use the unprecedented wavelength coverage and sensitivity of the IRS to explore the mid-infrared spectra of these stars. We find that two of the sample observed thus far, TW Hya and Hen 3-600, have both mid-infrared excesses with abrupt short-wavelength edges and crystalline silicate features which, taken together, demonstrate the processing of dust and the development of radial and vertical structure that has taken place in these 10 Myr-old disks.

## 2. Observations

We observed seven targets in the TW Hya association on 2004 January 4–5, using the Spitzer Space Telescope (Werner et al. 2004) and its Infrared Spectrograph (IRS; Houck et al. 2004). A journal of the observations appears in Table 1. For all but the two brightest objects, TW Hya and Hen 3-600, we operated the observatory in IRS Staring Mode, first performing a high-accuracy pointing peak-up, using the onboard Pointing Control Reference Sensor (PCRS), followed by exposures on the individual objects in both orders of the IRS Short Low spectrograph (henceforth SL; 5.3-15 $\mu$m, $\lambda/\Delta\lambda \sim 90$), and in some cases the second order of the IRS Long Low (LL2, 15–20 $\mu$m, $\lambda/\Delta\lambda \sim 90$) spectrograph. Each object was observed twice in each spectrograph and order, with the telescope nodded by one-third of the slit length in the spatial direction between consecutive exposures. TW Hya and Hen 3-600 were observed using IRS Spectral Mapping Mode with the Short Low and Short High (SH, 10–20 $\mu$m, $\lambda/\Delta\lambda \sim 600$) spectrographs. Omitting the pointing peakup, we performed a 2×3 raster (spatial × dispersion) centered on the star; the slit positions were separated by half (for SH) or three-quarters (SL) of their widths in the dispersion

---

[1] The IRS was a collaborative venture between Cornell University and Ball Aerospace Corporation funded by NASA through the Jet Propulsion Laboratory and the Ames Research Center.

direction, and by a third of their length in the spatial direction.

We reduced our spectra with the IRS team's Spectral Modeling, Analysis, and Reduction Tool (SMART; Higdon et al. 2004). We started with intermediate products from the Spitzer Science Center's IRS data-reduction pipeline that lacked only stray light and flatfielding corrections. From the low-spectral-resolution data, we extracted point-source spectra for each nod position using a variable width column which scaled with that of the instrumental point-spread function. Full-slit extractions were performed on the data from SH. We also made similar spectral extractions for observations of two photometric standard stars, $\alpha$ Lac (A1 V) and $\xi$ Dra (K2 III). We then divided each target spectrum, nod position by nod position, by the spectrum of one of the standard stars and multiplied by the standard's template spectrum (Cohen et al. 2003). The final spectra are the averages of the nod positions. This procedure gives good sensitivity and spectrophotometric uncertainty which we estimate to be 10%. It is important to distinguish this absolute photometric accuracy from the point-to-point fluctuation in the spectra, which is a measure of the accuracy of our spectral feature strengths and therefore of ability to identify spectral features and compare them to our models. The accuracy of our spectral feature identifications and feature strengths is limited by the point-to-point scatter (currently much larger than system noise, in bright objects) that is visible in the spectra. All features that significantly exceed the fluctuations, and are wider than the two-pixel IRS spectral resolution element, are significant. We expect soon to achieve 5% spectrophotometric accuracy and better spectroscopic accuracy with the same data, using flat fields acquired over the first six months of the mission.

## 3. Analysis and Conclusions

Of the seven systems we observed, just two, TW Hya and Hen 3-600, have infrared emission in excess of their photospheres within the IRS spectral range, the same fraction as obtained for the older systems studied with IRS by Jura et al. (2004). The two cases of mid-infrared excess appear in Figure 1. The spectrum of Hen 3-600 is the sum of that of Hen 3-600A (which is a spectroscopic binary; see e.g. Torres et al. 2003) and its nearby fainter companion Hen 3-600B (1.6″ radial separation); they were not resolved in either the SL or SH observations (3.6″ and 4.7″ diameter beams, respectively). Also displayed is the spectrum of TWA 6, representative of the rest of the sample in which we observe no excesses over their photospheres. Table 1 gives the 10 $\mu$m fluxes of all of the observed sources. The flux densities we observe for TW Hya are within 10% of those observed in Band 3 (5.8 $\mu$m) and Band 4 (8.0 $\mu$m) of the Spitzer Infrared Array Camera (Hartmann 2004).

Both of the systems with mid-infrared excesses show strong spectral features from silicates at 10 $\mu$m, indicating that the TWA members are indeed different from the older objects studied in IRS spectra by Jura et al. (2004), despite similar frequency of occurrence of measurable excess; smaller dust grains are much more abundant in the TWA objects and dusty material lies closer to the stars than in main-sequence stars with mid-infrared excesses. At wavelengths shortward of the silicate complex, the spectra of both TW Hya and Hen 3-600 quickly approach the photospheric emission of the primaries in their systems (K7 and M3, respectively); the spectra clearly show that there is essentially no emission from the disk at wavelengths $\lambda < 7\mu$m, as indicated by the excellent agreement between the 5–7 $\mu$m spectra of TW Hya and TWA 6, with similar spectral types. At wavelengths longer than the silicates, the complex gives way to continuum emission much brighter than the photosphere. This confirms in considerable detail the finding by Calvet et al. (2002), based upon ground-based broad-band photometry and modeling, of a *deficit* of emission from the inner few AU of these disks. Moreover, our observations confirm the existence of a flux deficit below 10 $\mu$m and substantial emission at longer wavelengths in the SED of Hen 3-600 (Jayawardhana et al. 1999a); these ground-based observations indicate that this excess originates in Hen 3-600A, though both A and B lie within the IRS slit.

Crystalline silicate features and indications of grain growth appear in the spectra of Hen 3-600 and TW Hya. The spectrum of Hen 3-600 is particularly rich. We observe several features in the 9–12.5 $\mu$m range observed by Honda et al. (2003b) and ascribed to a mixtures of forsterite (especially the 11.3 $\mu$m feature), enstatite and silica (9 and 12.5 $\mu$m), identified in Figure 1. In addition, we have detected features at 11.9, 13.7 and 16.1 $\mu$m. The 9-13.7 $\mu$m features are in the SL and SH wavelength overlap region and indeed, they are observed in the spectra of both modules. The 16 $\mu$m feature is identified with forsterite and has previously been detected in young stars and comets (e.g., Malfait et al. 1998). It is possible to explain part of the narrow 11.9 $\mu$m feature with forsterite or maybe even enstatite, but not the 13.7 $\mu$m feature. It is also possible that both of these features are produced by Mg-poor spinel (Fabian et al. 2001).The spectral features at 5–7 $\mu$m are an accurate match for photospheric features observed in the M3.5 V star GJ 1001A (Roellig et al. 2004). The high-resolution spectrum of TW Hya also reveals the presence of the strongest forsterite feature at 11.3 $\mu$m, and a 10 $\mu$m complex width consistent with that of Hen 3-600.

To give a more quantitative assessment of the degree of grain growth and crystallinity, we have taken from the model by Calvet et al. (2002) only the ingredients most relevant to the IRS wavelength range, but we have considered in more detail the mineralogical composition of the silicates. In this approximation, the disk emission consists of (1) radiation of the "wall" at the disk truncation radius $R_t$ illuminated frontally by the star with luminosity $L_*$. This emission, which dominates at long wavelengths, is modeled as a black body with temperature $T_{bb}$ and solid angle $\Omega_{bb}$; (2) radiation from the inner optically thin disk, of optical depth $\tau_i$, which accounts for the silicate emission (cf. (Calvet et al. 2002)). Fit to the spectra yields $T_{bb}$, $\Omega_{bb}$, $\tau$. We considered the following dust species in our model: glassy olivine, crystalline forsterite, crystalline ortho-enstatite, and crystalline quartz. To sample for dust size, we calculated mass absorption coefficients for the amorphous silicates using Mie theory and three size distributions, 1.9–2.1 $\mu$m, 0.09–0.2 $\mu$m, and 0.005–0.25 $\mu$m. Optical constants were taken from Dorschner et al. (1995) and Wenrich & Christensen (1996). For the crystalline silicates, we used the mass absorption coefficients from Koike et al. (1999). The density assumed for all silicates was 3.3 g cm$^{-3}$. We calculated the optical efficiency of the quartz component with a CDE model (Bohren & Huffman 1983). We took the grain size to be $a = 0.1\mu$m, and the range of aspect to be $b = 0.1a$–$10a$. The result was scaled arbitrarily in the fit.

The best fit structural parameters for TW Hya are $T_{bb} = 129$K, $\Omega_{bb} = 6.5 \times 10^{-14}$, and $\tau(R_t) \sim 0.2$ (at 10 $\mu$m). Taking the height of the wall or characteristic width (which is more appropriate for the nearly face-on TW Hya) as 0.10 times its radius, we obtain a truncation radius $R_t \sim 3.3$ AU, consistent with (Calvet et al. 2002). The best fit for Hen 3-600 is obtained for $T_{bb} = 200$ K, $\tau(R_t) \sim 0.1$, and $\Omega_{bb} = 8.3 \times 10^{-15}$, which implies $R_t \sim 1.3$ AU.

Our best model for TW Hya implies a mass fraction of amorphous silicates of 0.007, with 70% in grains of 2 $\mu$m size, indicating substantial grain growth from ISM mixtures (with maximum grain sizes of $\sim 0.2$ $\mu$m). On the other hand, the width of the 10 $\mu$m feature and the slope between 10 and 20 $\mu$m require some silicate in crystalline form, with a mass fraction of $3.5 \times 10^{-4}$ in forsterite and $4.4 \times 10^{-3}$ in ortho-enstatite. To fit the spectrum of Hen 3-600 we have added silica to our dust mixture, following Honda et al. (2003b). We require a mass fraction of amorphous silicates of 0.006, similar to that of TW Hya, although with comparable fractions of small and large grains. On the other hand, the mass fractions of forsterite and ortho-enstatite are $3.3 \times 10^{-3}$ and $3.5 \times 10^{-3}$, an order of magnitude higher than those in TW Hya. The comparison of these models with observation is shown in Figure 2. Both systems still emit more than the model at wavelengths around 8 $\mu$m; in TW Hya the model is slightly fainter than the data around 13 $\mu$m. This deficit could indicate the presence of another warm component in the interior of the disk (distributed or compact), or the absence in the model of important ingredients of the dust grains, or their correct mixture. We will address this question with more accurate and comprehensive modeling in the future.

The degree of grain growth and crystallinity found in the TWA stars is consistent with those in the most evolved among the brightest Herbig Ae/Be stars studied with ISO (Bouwman et al. 2001, 2003). In particular, the presence of silica in Hen 3-600, with the higher fraction of crystals, is consistent with the correlation between the abundance of SiO$_2$ and increasing crystallinity found by Bouwman et al. (2001). The SEDs of these two objects indicates significant dust clearing in the inner disk. If larger bodies have been formed in TW Hya and Hen 3-600, and are responsible for the inner clearing of the disk, it seems plausible to speculate that collisions of such bodies might produce smaller particles which are heated to temperatures necessary to produce more crystalline silicate dust. Further studies of pre-main sequence objects with IRS will help confirm whether this difference between cleared and optically thick pre-main sequence disks is common.

We are grateful to our anonymous referee for improving our paper through an extremely prompt, thorough, and thoughtful review. We also thank Tom Roellig for access to the spectrum of GJ 1001A. This work is based on observations made with the Spitzer Space Telescope, which is operated by the Jet Propulsion Laboratory, California Institute of Technology under NASA contract 1407. Support for this work was provided by NASA through Contract Number 1257184 issued by JPL/Caltech and through the Spitzer Fellowship Program, under award 011 808-001.

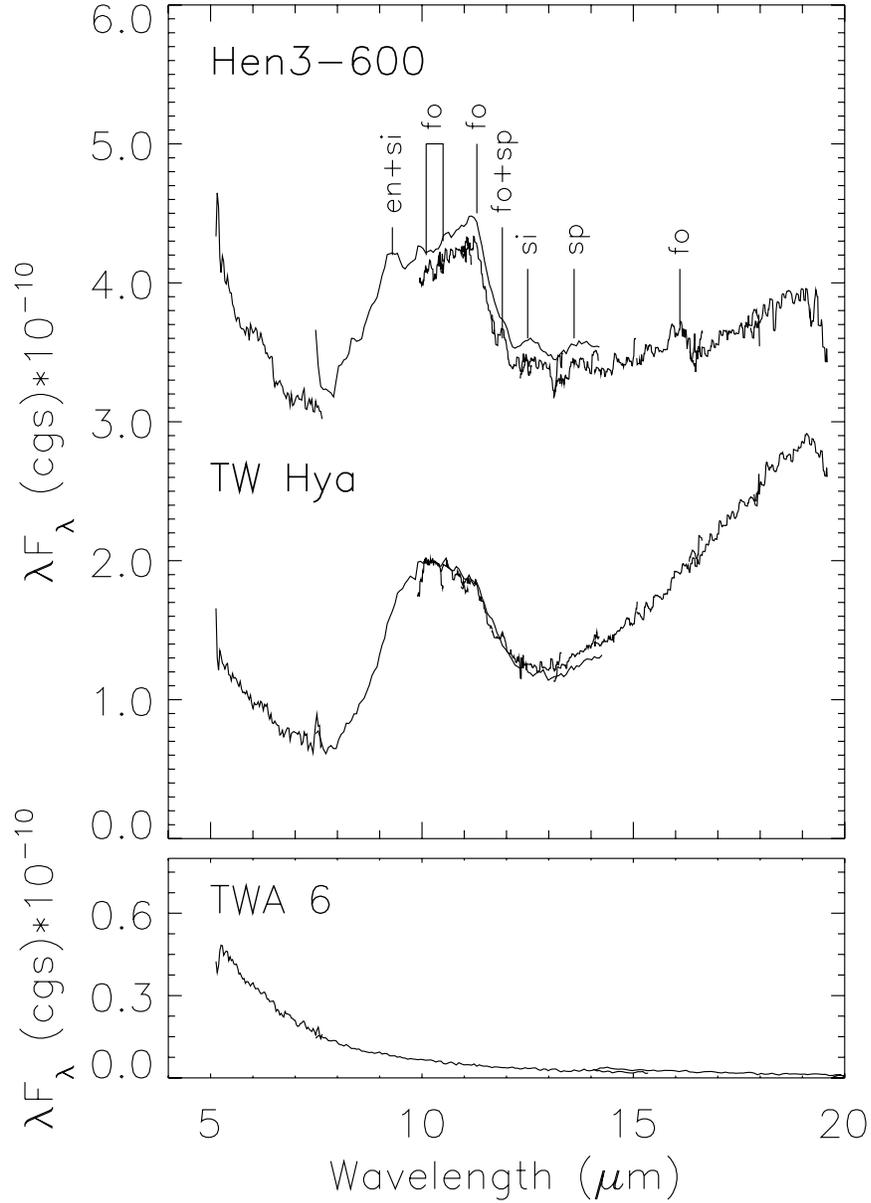

Fig. 1.— IRS spectra of TW Hya, Hen 3-600, and TWA 6. The SH spectra of TW Hya and Hen 3-600 have been median smoothed in 0.075 $\mu$m wavelength bins. Hen 3-600 has been offset by +1.5 along the y-axis for display purposes. As discussed in the text, the identities of the various features are as follows: "fo" = forsterite, "en" = enstatite, "si" = silica, and "sp" = spinel. The TW Hya and Hen 3-600 spectra shown are composites of all orders of SL and SH, and TWA 6 of SL and LL2. No adjustments were made between the overlapping spectral segments from different orders or spectrograph modules for a given target; the small offsets visible between SL1 and SH segments serves to illustrate the magnitude of the spectrophotometric uncertainties. Artifacts of poor illumination at the extremes of orders (e.g. at 7.5 and 19 $\mu$m) also remain in the spectra.

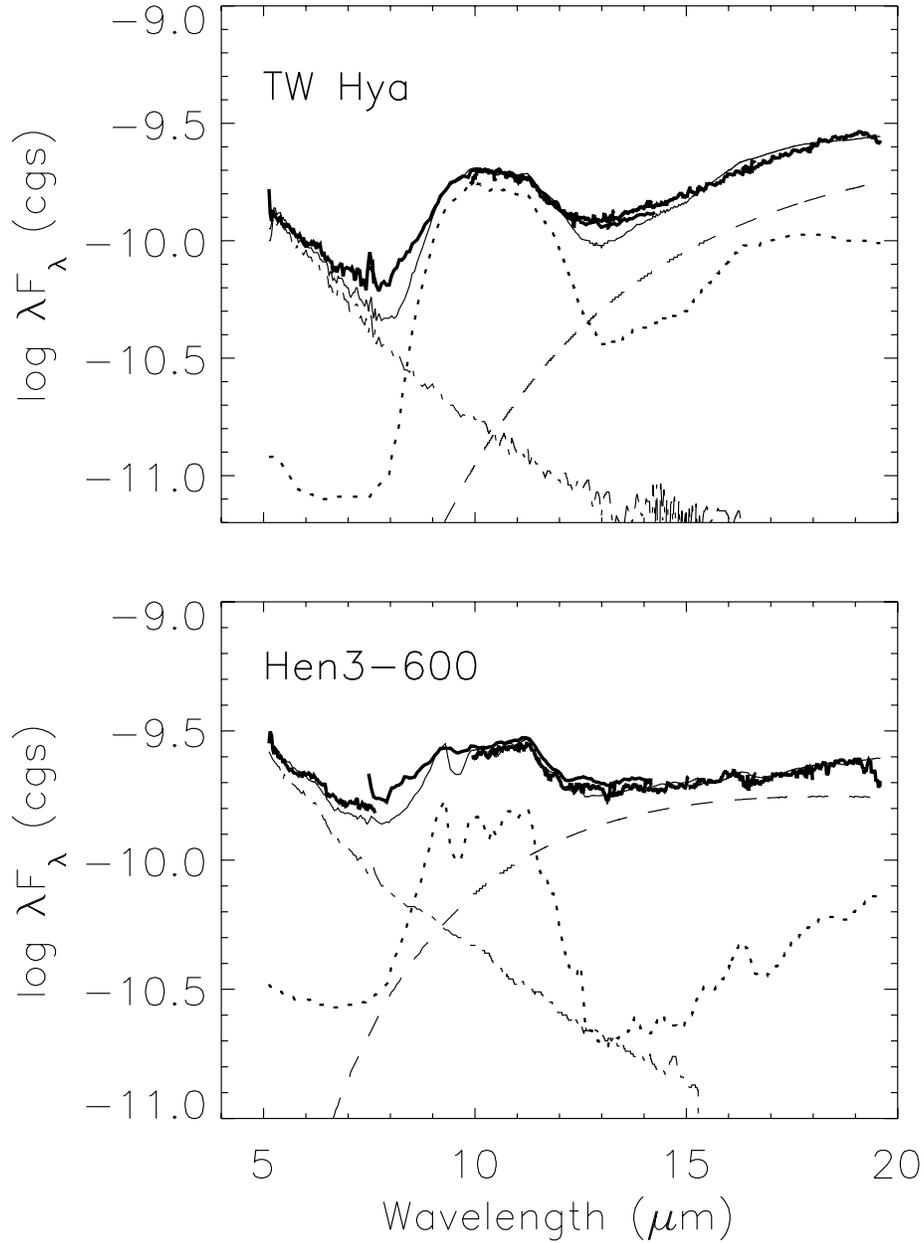

Fig. 2.— Top panel: Spectrum of TW Hya (heavy solid lines) compared to model fluxes (light solid lines). The contributions for the different regions of the disk are shown: the wall emission (dashed line) and the optically thin inner region (dotted line). The spectrum of TWA6, scaled in proportion to the K magnitude and by an additional factor of 1.5, is taken as the stellar photosphere (dot-dash line). Lower panel: Spectrum of Hen 3-600 compared to model fluxes. The same line styles as in the top figure are used to represent the model components. The spectrum of the M3 star GJ 1001A, scaled to the K magnitude of Hen 3-600 and by an additional factor of 1.9, is taken as the photosphere (dot-dash line).

– 12 –Table 1.

| Source | RA (J2000) | Dec (J2000) | Spectral Type | Obs Mode | Modules | IRS 10 $\mu$m flux density (mJy) |
|---|---|---|---|---|---|---|
| TW Hya | 11:01:51.91 | -34:42:17 | K7e | 2x3 Map | SL, SH, LH | 624 (SH), 665 (SL) |
| TWA 2[1] | 11:09:13.90 | -30:01:39 | M2e+M2 (A+B) | Staring | SL, LL | 71 |
| Hen3-600[1] | 11:10:28.00 | -37:31:53 | M3e+M3.5 (A+B) | 2x3 Map | SL, SH, LH | 830 (SH), 917 (SL) |
| TWA 5[1] | 11:31:55.40 | -34:36:27 | M1.5+M8.5 (A+B) | Staring | SL, LL | 61 |
| TWA 6 | 10:18:28.80 | -31:50:02 | K7 | Staring | SL, LL | 22 |
| TWA 9A | 11:48:24.22 | -37:28:49 | K5 | Staring | SL, LL | 29 |
| TWA 9B | 11:48:23.70 | -37:28:49 | M1 | Staring | SL, LL | 11 |

[1]TWA 2, Hen 3-600 and TWA 5 are multiple star systems, each with a nearby B component (or more) which are unresolved by the IRS observations, regardless of the module. The positions observed are of the primary A components.